\documentclass[conference]{IEEEtran}
\IEEEoverridecommandlockouts

\usepackage{cite}
\usepackage{amsmath,amssymb,amsfonts}
\usepackage{algorithmic}
\usepackage{graphicx}
\usepackage{textcomp}
\usepackage{xcolor}
\usepackage{multirow}
\usepackage{tabularx,array}
\usepackage{hyperref}
\usepackage{booktabs}
\usepackage{xcolor}

\def\BibTeX{{\rm B\kern-.05em{\sc i\kern-.025em b}\kern-.08em
    T\kern-.1667em\lower.7ex\hbox{E}\kern-.125emX}}
\begin{document}

\title{Fault Injection in OpenAPI Specifications for Evaluating Black-Box Testing Effectiveness}

\author{\IEEEauthorblockN{Hamza Bin Mazhar, Yuqing Wang, Mika Mäntylä}
\IEEEauthorblockA{\textit{Department of Computer Science, University of Helsinki}, Helsinki, Finland \\
\{hamza.mazhar, yuqing.wang, mika.mantyla\}@helsinki.fi}
\thanks{© 2026 IEEE. Personal use of this material is permitted. Permission from IEEE must be obtained for all other uses, in any current or future media, including reprinting/republishing this material for advertising or promotional purposes, creating new collective works, for resale or redistribution to servers or lists, or reuse of any copyrighted component of this work in other works.}
}

\maketitle

\begin{abstract}
OpenAPI specifications are the primary input for black-box testing tools in microservice systems (MSS), yet prior work shows these specifications are often incomplete, inconsistent, or incorrect. Despite this, most studies on OpenAPI-based black-box testing assume correct specifications and evaluate tool performance. We address this gap by introducing a literature-grounded taxonomy of six OpenAPI specification fault classes. We inject faults at five severity levels, and evaluate the resulting mutated specifications on two microservice benchmarks, TrainTicket and SocialNetwork, using three testing tools: EvoMaster, RESTler, and Schemathesis. We measure the impact of these faults using code coverage, specification coverage, request/response quality, and behavioral diversity. Our results show that specification faults cause strong and heterogeneous degradation patterns across testing tools and systems. Faults in method semantics cause broad degradation across all metrics, while others, such as modifications to response codes, remain weak. Relaxations of schema constraints cause hidden degradation, with no impact on code and specification coverage but a large impact on request/response quality. These findings demonstrate that specification quality directly shapes black-box API testing effectiveness. Also, code and specification coverage-only evaluations can understate the impact of specification faults on black-box testing in MSS and should be complemented by request/response quality and behavioral diversity.
\end{abstract}


\section{Introduction}
Many microservice systems (MSS) expose REST-based interfaces, especially at system boundaries, enabling black-box testing as a viable and scalable testing approach. OpenAPI specifications~\cite{openapi2026} serve as the primary source of information for black-box testing by providing the testing tools~\cite{arcuri2019, atlidakis2019, hatfield-dodds2022} with descriptions of the available operations, input parameters and expected responses to generate and assess test cases. However, this reliance raises a concern. 

Prior work has shown that the OpenAPI specifications are commonly plagued with faults~\cite{martin-lopez2021, arcuri2024, kimautoresttest}, inconsistencies and missing information~\cite{eddoubi2018, martin-lopez2019, neumann2021}. If these flawed specifications are utilized for the testing process, the process could be less effective or incomplete.

This paper studies that problem directly. Instead of treating faulty specifications as random noise, we use fault injection to systematically mutate OpenAPI specifications. We develop a literature-grounded taxonomy for OpenAPI specification faults and inject these faults at several severity levels into the baseline specifications, with no modifications made to the underlying services implementations. We utilize the testing tools as a mechanism to evaluate the impact of the faults in the specifications. 

We evaluate this design utilizing two benchmark MSS, TrainTicket~\cite{TrainTicket} and SocialNetwork~\cite{SocialNetwork}, and three popular OpenAPI-based black-box testing tools, namely EvoMaster~\cite{arcuri2019}, RESTler~\cite{atlidakis2019}, and Schemathesis~\cite{hatfield-dodds2022}. 
We study two systems rather than many smaller APIs because our experimental unit is the specification variant, not the system itself. Prior tool comparisons~\cite{kim2022, corradini2021empiricalcomparisonblackboxtest, zhang2023, 10.1145/3765744}, alongside related empirical evaluations~\cite{sahin2025wfcwfdwebfuzzingcommons, martin-lopez2021} evaluate many single-service APIs under fixed specifications, whereas our design mutates specifications across multiple fault classes and severity levels, producing over 50 mutated specification variants and more than 450 runs across two microservice benchmarks of contrasting size, the large TrainTicket and the small SocialNetwork.

We assess the variation in code coverage, specification coverage, API request/response quality, and server-side behavioral diversity as a function of both the fault class and the degree of severity of each fault. The goal of this study is to illustrate how faults in the specifications could lead to failures or degraded performance in OpenAPI-based black-box testing tools. Fault injection illustrates why microservice testing should consider specification quality. The paper makes four main contributions.

\begin{enumerate}
    \item We define a taxonomy of OpenAPI fault classes grounded in previous literature. 
    \item We inject faults into specifications at various degrees of severity.
    \item We conduct an empirical study across two microservice benchmark systems and three OpenAPI-based black-box testing tools to evaluate the extent to which specification faults impact testing outcomes.
    \item We identify which of the injected fault classes result in the greatest degradation of observed testing behavior.
\end{enumerate}

\section{Background}
\label{sec:Background}
\subsection{OpenAPI Specification Faults}
Specification quality directly affects testing effectiveness. Specification faults
can cause tools to generate incomplete, invalid, or no requests at all~\cite{kim2022}, and empirical evaluations show that tools may fail to infer callable endpoints or even crash and not make any HTTP calls due to schema faults~\cite{sahin2025wfcwfdwebfuzzingcommons}. Another paper~\cite{martin-lopez2021} notes that specifications may not be suitable to generate realistic test inputs, or they may simply be wrong. The authors report having to manually alter the specifications to compensate for missing input constraints.

Many empirical studies have reported that specification faults are common in practice. For example, Ed-douibi et al.~\cite{eddoubi2018} tested 91 OpenAPI specifications from APIs.guru and found that 40\% produced errors when a correct client calls a real API. The errors are split between faults in the specifications themselves and the implementation. They attribute a large share to missing or incorrect constraints and response schema mismatches. Their method is observational and they do not control which faults are present in the specifications. 

Martin-Lopez et al.~\cite{martin-lopez2019} cataloged inter-parameter dependencies in 40 real-world APIs and found that four out of five APIs contain dependencies that OpenAPI cannot formally express. Neumann et al.~\cite{neumann2021} analyzed 500 public APIs and found that, while HTTP verb semantics are followed by services, 29.8\% of the services omit error message explanations, which contributes to incomplete response documentation. Kim et al.~\cite{kim2022} reported that almost 99\% of the 1,000+ specifications they checked on APIs.guru contained some mismatch.

\subsection{Benchmarking Black-Box Testing Tools}

Several OpenAPI-based black-box testing tools for APIs have been proposed in literature, including EvoMaster~\cite{arcuri2019}, RESTler~\cite{atlidakis2019}, Schemathesis~\cite{hatfield-dodds2022}. Studies have then performed empirical comparisons of subsets of these tools under fixed specifications. Kim et al.~\cite{kim2022} compare 10 such tools on 20 open source services. Sahin et al.~\cite{sahin2025wfcwfdwebfuzzingcommons} extend this comparison to include a larger benchmark. Other studies have also compared or discussed black-box API testing tools in different contexts, such as comparative evaluations~\cite{martin-lopez2021,corradini2021empiricalcomparisonblackboxtest,zhang2023}, industrial case studies on evolving applications~\cite{10.1145/3765744} and industrial adoption in service-oriented environments~\cite{Poth2024}. 
 
However, none of these studies examine the impact of faults in OpenAPI specifications on the effectiveness of black-box API testing for MSS. Furthermore, all previous evaluations assume that the specifications, used for each execution of the black-box testing tools, are fixed. This implies that the performance differences among the various tools are attributed to the strategy employed by each of the testing tools and not to the potential faults in the specifications.

\subsection{Specification Enrichment}\label{subsec:SpecificationEnrichment}
Several approaches compensate for under-specified schemas, by recovering missing parameters through white-box analysis or enriching specifications using NLP, LLMs, or runtime feedback to infer missing constraints~\cite{kim2023nlptorest,Jain2025,arcuri2021}. However, they aim to compensate for the shortcomings of the specification by some technical solution. They do not study how testing behavior degrades when faults remain present in the specification, which is the focus of this paper.

In addition to faulty specifications, implicit inter-parameter dependencies are another factor that hampers automated test generation. Since these dependencies cannot be explicitly described in OpenAPI specifications~\cite{martin-lopez2019}, approaches have been developed for extracting them automatically, such as NLPtoREST~\cite{kim2023nlptorest} and LlamaRestTest~\cite{llamaresttest}, which collect additional data to compensate for missing dependencies and constraints.

\subsection{Mutation Testing as Methodology} Prior mutation studies most similar to ours come from the earlier SOAP/WSDL Web-services literature. These studies perform mutations directly in WSDL descriptions. Adjacent studies consider contract mutation and SOAP-message mutation for Web-service testing~\cite{jiang2009contract,chen2014soapmutation,bluemke2016mutant}. However, these studies were proposed for mutation-based test generation or mutation-based test assessment, and not to investigate how specification faults would affect the effectiveness of black-box testing. 
 
\subsection{Our motivation}
Prior studies indicate that OpenAPI-based black-box testing tools depend significantly on the quality of the OpenAPI specifications, and that specifications in real-world scenarios are frequently either incomplete or incorrect. Such faults can reduce the coverage generated by a testing tool, generate invalid requests, or even cause the testing tools to crash. What is lacking is a controlled study that introduces specific OpenAPI faults at different severity, and examines how strongly black-box testing is affected by those faults in MSS. 

\section{Fault taxonomy}
\label{sec:Fault taxonomies and mutation-testing background} 
We derived the fault taxonomy from prior studies on OpenAPI specifications and OpenAPI-based API 
testing. We organize these faults into six classes based on how the specification supports black-box API testing: over-constrained requests, under-constrained requests, semantic violations in API designs, oracle limitations, structural specification weaknesses, and composite faults. Table~\ref{tab:faultclass-grounding} summarizes the mapping between prior work and the resulting fault classes. We discuss each category in detail below. 

\textbf{Constraint-related faults} focus on incompleteness and inconsistency in the OpenAPI request and schema constraints. Previous studies report missing constraints~\cite{martin-lopez2019, 10.1109/ICSE-SEIP52600.2021.00016}, schema mismatches, underspecified schemas, and other forms of specification incompleteness are present~\cite{kim2022, zhang2023, hatfield-dodds2022, 9425926}. These studies motivate our \texttt{REQUEST\_STRICT} and \texttt{REQUEST\_LOOSE} classes. In the mutation design process, the former tightens constraints, e.g. \texttt{required}, \texttt{minLength}, and numerical bounds, while the latter systematically loosens or removes such constraints. Thus, these two fault classes represent two opposite directions of contract request strength perturbation. These classes let us measure how sensitive testing tool outputs are to specification side constraint quality. 

\textbf{Semantic faults} class is about violations of API design rules, particularly with regard to HTTP-method usage. Studies on REST patterns and antipatterns~\cite{10.1007/978-3-662-45391-9_16, 10.1007/978-3-319-46295-0_10} and on compliance with HTTP best practices~\cite{neumann2021, 10.1007/978-3-319-38791-8_2, 10.1007/s10664-023-10367-y} motivate the \texttt{METHOD\_SEM} class. These studies consistently treat HTTP verb semantics as a meaningful and relatively stable convention, which makes semantic drift in method usage a realistic specification fault. In our fault injection, this class remaps methods on a specified set of operations (\texttt{GET}\(\leftrightarrow\)\texttt{POST}, \texttt{PUT}\(\leftrightarrow\)\texttt{PATCH}, \texttt{DELETE}\(\rightarrow\)\texttt{GET}) while keeping all paths and implementation unchanged.

\textbf{Oracle faults} addresses the oracle problem in API testing and the dependence of test evaluation on response documentation. Work on specification-based oracles~\cite{6963470, eddoubi2018, 10.1145/3597926.3598114, CorradiniAutomated2022}, metamorphic testing~\cite{8074764}, empirical comparison of specification-based API testing tools~\cite{corradini2021empiricalcomparisonblackboxtest, Golmohammadi2023} motivates the \texttt{RESPONSE\_ORACLE} class. These studies demonstrate that OpenAPI-based API testing relies on documented status codes and response schemas as practical response-side oracles, and that incomplete, weak, or mismatched response descriptions can weaken validation, misclassify behavior, or leave incorrect responses undetected. In our fault injection, this class removes some documented \texttt{500} responses and tightens some documented \texttt{2xx} response schemas.

\textbf{Structural faults} target the parseability and internal consistency of the specification document. Studies demonstrate that testing tools are generally fragile when specifications are structurally invalid, excessively complex, or improperly constructed. Empirical studies of tool failures on actual APIs~\cite{kim2022, corradini2021empiricalcomparisonblackboxtest, sahin2025wfcwfdwebfuzzingcommons} and structural deficiencies in OpenAPI artifacts~\cite{hatfield-dodds2022, 9779839} motivate the \texttt{PARSE\_BROKEN\_REF} class. From these studies, we derive the notion that faults in the structural integrity or tool-compatibility of the specification itself prevent meaningful testing before any type of exploration begins. In our fault injection, this is implemented as a single internal \texttt{\$ref} corruption. OpenAPI specifications use \texttt{\$ref} pointers to reference reusable schema definitions somewhere else in the document. If we corrupt such a reference to point to a nonexistent definition, it forces the testing tool to either fail at parsing or skip the affected schema.

\textbf{Composite faults} come from mutation testing and fault interaction research. Studies on higher-order mutation~\cite{JIA20091379, 10.1145/2642937.2643008}, composite faults~\cite{7927962}, and fault interactions~\cite{10.1145/125489.125473} motivate the \texttt{COMPOSITE} class. Studies on REST antipatterns~\cite{10.1007/978-3-662-48616-0_11,10.1007/978-3-662-45391-9_16} demonstrate that real-world APIs present coexisting multiple design faults. These studies support the view that combining multiple specification faults is both theoretically meaningful and practically realistic. In our fault injection, this class applies a deterministic split of a fixed severity budget across \texttt{METHOD\_SEM}, \texttt{REQUEST\_STRICT}, \texttt{REQUEST\_LOOSE}, and \texttt{RESPONSE\_ORACLE} on disjoint sets of selected operations.

\begin{table*}[t]
\caption{Fault Taxonomy Mapping with Prior Work}
\label{tab:faultclass-grounding}
\centering
\renewcommand{\arraystretch}{1.3}
\setlength{\tabcolsep}{6pt}
\footnotesize
\begin{tabular}{@{}p{2.5cm}p{2cm}p{6cm}p{6.36cm}@{}}
\toprule
\textbf{Fault Class} & \textbf{Source} & \textbf{Description} & \textbf{Example} \\
\midrule
\texttt{REQUEST\_STRICT}
& \cite{bluemke2016mutant,martin-lopez2019,10.1109/ICSE-SEIP52600.2021.00016,kim2022,marculescufaults2022}
& Tighten request-side contracts beyond intended behavior to alter request patterns and downstream metrics when tools consume stricter constraints.
& Make an optional parameter required; change \texttt{\{"type":"integer"\}} to \texttt{\{"type":"integer",} \texttt{"minimum":10\}}; narrow an enum or regex. \\
\hline
\texttt{REQUEST\_LOOSE}
& \cite{bluemke2016mutant,martin-lopez2019,10.1109/ICSE-SEIP52600.2021.00016,zhang2023,hatfield-dodds2022,marculescufaults2022,kim2022,CorradiniAutomated2022}
& Relax or remove schema constraints, making the specification too permissive and helping test whether stricter schema contracts improve test generation.
& Change \texttt{required: true} to \texttt{false}; drop \texttt{enum}, \texttt{pattern}, \texttt{format}, or \texttt{bounds}. \\
\hline
\texttt{METHOD\_SEM}
& \cite{neumann2021,10.1007/978-3-319-38791-8_2,10.1007/978-3-319-46295-0_10,10.1007/s10664-023-10367-y,10.1007/978-3-030-87568-8_10,10.1007/978-3-030-91431-8_10}
& Remap HTTP methods to disturb operation semantics and shift request quality and coverage-style metrics if tools rely on method semantics.
& Replace \texttt{GET} with \texttt{POST}; swap \texttt{PUT} and \texttt{PATCH}; map \texttt{DELETE} to \texttt{GET}. \\
\hline
\texttt{RESPONSE\_ORACLE}
& \cite{6963470,8074764,10.1145/3597926.3598114,CorradiniAutomated2022,eddoubi2018,corradini2021empiricalcomparisonblackboxtest,Golmohammadi2023,kim2022,neumann2021}
& Mutate response documentation while keeping the tool exploration side separate so specification-based oracle checks become weaker or misleading.
& Remove documented \texttt{500} responses; tighten a \texttt{2xx} response schema by composing it with an additional required marker property. \\
\hline
\texttt{PARSE\_BROKEN\_REF}
& \cite{hatfield-dodds2022,sahin2025wfcwfdwebfuzzingcommons,9779839,corradini2021empiricalcomparisonblackboxtest,kim2022,Golmohammadi2023}
& Corrupt an internal \texttt{\$ref} to a missing definition to create a hard specification-usability boundary and cause parse/compilation/run failure or collapse.
& Change \texttt{\$ref:} \texttt{\#/components/schemas/User} to \texttt{\$ref:} \texttt{\#/components/schemas/\_\_missing\_\_}. \\
\hline
\texttt{COMPOSITE}
& \cite{7927962,JIA20091379,10.1145/2642937.2643008,10.1145/125489.125473,10.1007/978-3-662-48616-0_11,10.1007/978-3-662-45391-9_16}
& Apply multiple fault classes within the same variant specification to measure the accumulated effect.
& --- \\
\bottomrule
\end{tabular}
\end{table*}

\section{Empirical Study Design}
\label{sec:Empirical Study Design}
Using the fault taxonomy from Section~\ref{sec:Fault taxonomies and mutation-testing background}, we empirically evaluate the impact of systematically injected faults in baseline OpenAPI specifications on REST API testing tools. 
To guide this study, we first define three research questions, and present our experimental design in the following subsections. 
\begin{itemize}
    \item \textbf{RQ1:} What is the impact of the fault classes in each system and tool combination, and what degradation patterns do they produce?
    \item \textbf{RQ2:} Which degradation effects are shared across tools, and which are tool-specific?
    \item \textbf{RQ3:} Which metric families capture the degradation signal of each fault class?
\end{itemize}

\subsection{Subject Systems}
The systems under test (SUT) for these fault injection experiments are SocialNetwork~\cite{SocialNetwork} and TrainTicket~\cite{TrainTicket}. SocialNetwork contains 21 internal services exposed through 6 gateway-level operations. TrainTicket is a train ticket booking MSS consisting of 44 microservices. We treat TrainTicket as our primary system and SocialNetwork as a smaller confirmatory case that tests whether the same fault-effect patterns appear at a much smaller scale. 

We use two benchmark MSS rather than a larger set of smaller services for three reasons. \textit{First}, each specification variant requires a full MSS deployment with reset isolation across three tools and three seeds. Across both systems, we produce over 50 specification variants, resulting in over 450 runs and approximately 88 hours of execution time on dedicated infrastructure. \textit{Second}, our study targets a different empirical objective than prior tool comparisons. Those studies evaluate tools across many APIs under fixed specifications~\cite{kim2022, sahin2025wfcwfdwebfuzzingcommons, martin-lopez2021, corradini2021empiricalcomparisonblackboxtest, zhang2023}. Our experimental unit is not the system but the specification variant. \textit{Third}, both subject systems are heterogeneous  benchmark MSS with independently deployable services, inter-service communication and API gateways. SocialNetwork provides no OpenAPI specifications, so we constructed a baseline specification entirely from gateway and interface artifacts. This manual construction effort is itself part of the study cost and a key reason we do not scale to many systems. More broadly, internal service interfaces in heterogeneous MSS are not always described in OpenAPI, since they often use styles beyond HTTP/REST such as gRPC~\cite{grpc} or Thrift~\cite{apachethrift}, and may require manual construction.

\subsection{OpenAPI Specifications for SUTs}
\label{Sec:OpenAPI Specifications for SUTs}
\subsubsection{TrainTicket}
For TrainTicket, we built the baseline specifications by querying the /v2/api-docs endpoint exposed by Spring Boot services via the Springfox library~\cite{springfox}, which auto-generates OpenAPI specifications from annotated controllers at runtime. We retrieved specifications for 30 services from their live Kubernetes deployments. Four services did not return usable output from /v2/api-docs. For those services, we manually constructed the specifications using source-level inspection and runtime endpoint analysis. The TrainTicket baseline contains 34 services, 172 paths, and 211 operations.

\subsubsection{SocialNetwork}
SocialNetwork does not expose any OpenAPI endpoint per service. SocialNetwork is a Thrift-based system, so we constructed the baseline from static artifacts in the NGINX + Lua + Thrift stack. We extracted gateway routes and handler mappings from the NGINX configurations. We retained only the \texttt{/wrk2-api/} endpoints, as these are stateless, JSON-oriented benchmark interfaces suitable for OpenAPI-based testing. The session-oriented \texttt{/api/} routes return HTML responses and are not appropriate for this purpose. We then mapped the endpoints to Thrift service methods, parsed the \texttt{.thrift} and \texttt{gen-py} files to infer parameter types and request/response schemas, extracted status codes from Lua handlers, and assembled the final specification from these intermediate artifacts. The SocialNetwork baseline contains 5 services, 6 paths, and 6 operations.

These baselines may contain faults within themselves. We assess their correctness in two ways. First, they reflect the running services by construction, built by querying the exposed services, or otherwise from source, gateway, and handler inspection. Second, a perfect baseline is not required by design. Every metric used in the evaluation is a baseline-relative change. Imperfections already present in the baselines are held constant across all the mutated variants and cancel out, which isolates the effects of the injected faults and makes the comparison robust to manual construction errors.

\subsection{Fault Injection Design}
\label{Sec:Fault Injection Design}
We begin with the baseline OpenAPI specification for the two SUTs. From this baseline, we produce mutated specification variants using the fault classes (defined in Section~\ref{sec:Fault taxonomies and mutation-testing background}) and severity levels, which is the percentage of operations in the baseline specification that are mutated using the fault classes (P15: 15\%, P30:30\%, P45:45\%, P60:60\%, P75:75\%). For a given baseline specification with \(N\) operations and severity \(s\) (expressed as a fraction), the number of mutation actions to be applied \(A\) is: \(N \cdot s\), rounded to the nearest integer. In all these cases, the service implementation remains unchanged. The only thing changing is the specification. 

We generate one mutated specification per fault class at each severity level, so the number of variants per specification is fixed by the fault classes and severity levels and does not depend on the size of the system. The action count $A$ sets how many operations each variant mutates, so larger systems mutate more operations at the same severity rather than producing more variants. 
As we mutate a proportion of the specification rather than a fixed count, the severity level stays comparable across systems of very different size. Each variant carries a single fault class, so the effect of each fault type can be read in isolation. \texttt{COMPOSITE} is the only class that combines fault classes within one specification, which reflects the way real specification faults co-occur. As affected operations are chosen under the fixed seeds (Section~\ref{sec:Experimental Setup and Execution}), the full set of variants for a system can be reproduced from its baseline specification.

An injector script produces each mutated specification variant (fault class x severity) along with metadata indicating which operations were modified from the baseline specification. Table~\ref{tab:severity_budget} shows the calculations for both systems at all severity levels.
Four cases are treated differently:
\begin{enumerate}
    \item \texttt{PARSE\_BROKEN\_REF} is never graded by severity. It is a parseability check and always has a single mutation action. RESTler fails deterministically under this fault, while EvoMaster and Schemathesis still execute with warnings. This illustrates tool-level differences in specification tolerance.
    \item \texttt{COMPOSITE} uses the total budget \(A\), but divides it among the remaining four fault classes (except for \texttt{PARSE\_BROKEN\_REF}).
    \item For TrainTicket, \texttt{METHOD\_SEM} can provide fewer mutation actions, \(A\), than calculated, at higher severities. This happens because opposite method pairs (\texttt{GET\(\leftrightarrow\)POST}, \texttt{PUT\(\leftrightarrow\)PATCH}) selected on the same path can result in one effective remapping rather than two mutations. 
    \item Since SocialNetwork has only 6 operations, \texttt{COMPOSITE} is only injected at \(P60\) and \(P75\), where the severity is large enough to split meaningfully across four fault classes.
\end{enumerate}

\begin{table}[t]
\caption{Severity budgets for the two baseline specifications.}
\label{tab:severity_budget}
\begin{center}
\begin{tabular}{lcc}
\hline
\textbf{Severity} & \textbf{TrainTicket (\(N=211\))} & \textbf{SocialNetwork (\(N=6\))} \\
\hline
P15 & 32  & 1 \\
P30 & 63  & 2 \\
P45 & 95  & 3 \\
P60 & 127 & 4 \\
P75 & 158 & 5 \\
\hline
\end{tabular}
\end{center}
\vspace{-0.5cm}
\end{table}

\subsection{Testing Tools used}
\subsubsection{Rationale for selection}

We reviewed 83 papers on API testing and fuzzing that use OpenAPI specifications as the primary input artifact for black-box testing. After filtering for papers that describe automated testing tools, provide empirical comparisons or evaluations of such tools, we compiled a list of all tool names reported across these studies. From this list, we selected EvoMaster, RESTler, and Schemathesis as the three tools for our study, since prior comparative evaluations consistently identify them among the strongest and most effective REST API testing tools~\cite{kim2022,sahin2025wfcwfdwebfuzzingcommons,zhang2023,10.1145/3765744}. 

\subsubsection{Configuration of tools}
Before running our primary fault injection experiments, we configured each tool in its intended operating mode and performed a per-system budget calibration. 
We ran EvoMaster in black-box mode, and Schemathesis and RESTler in fuzzing mode. For each system, we evaluated the tools over a range of budget levels.
For this, we considered runtime-based budgets for EvoMaster and RESTler (1m, 5m, 10m, 30m, 1hr) and example-based budgets for Schemathesis (1ex, 5ex, 15ex, 100ex, 500ex, 1000ex).

We selected the final budgets at the point where Line coverage, Branch coverage, Schema Operation Coverage, Schema 2xx Coverage, and the traffic quality being exercised by the test had all stabilized. Beyond this point, we observed that additional budget increased the time cost without meaningful change to testing behavior. This calibration, combined with the fixed seed set described in Section~\ref{sec:Experimental Setup and Execution}, helps ensure that observable differences in the main experiments are attributed to specification faults rather than budget or randomness effects.

Thus, the final testing configurations for the two SUTs are as follows: (1) For both TrainTicket and SocialNetwork, we ran EvoMaster in black-box mode and RESTler in fuzzing mode, both with a total runtime of 10 minutes; (2) For the Schemathesis budget, we ran TrainTicket on 100 maximum examples, and SocialNetwork on 1000 maximum examples. 

\subsection{Experimental Setup and Execution}
\label{sec:Experimental Setup and Execution}
The fault injection runs for TrainTicket and SocialNetwork followed the same general protocol. We tested each system on five severity levels (\texttt{P15}, \texttt{P30}, \texttt{P45}, \texttt{P60}, \texttt{P75}). For each specification variant at each severity, we ran all three tools on a fixed seed set of three values. We used the same fixed seeds for both fault generation and tool execution.
This ensures reproducible runs and isolates specification faults from tool randomness. For both systems, we first collected baseline metrics for the three tools using the baseline specifications described in Section~\ref{Sec:OpenAPI Specifications for SUTs}.

For \textbf{TrainTicket}, each of the five severity levels contains six mutated specifications, one per fault class. Each severity level yields \(6\) runs per tool. Repeating these runs with three seeds gives \(6 \times 3 = 18\) runs per tool per severity level. Across all five severity levels, this amounts to \(18 \times 5 = 90\) runs per tool. Adding the baseline configuration, which was also executed once per seed, gives \(90 + 3 = 93\) runs per tool. Across the three tools, this results in \(93 \times 3 = 279\) runs in total.

For \textbf{SocialNetwork}, the \texttt{P15}, \texttt{P30}, and \texttt{P45} levels have five mutated specifications, per fault class except \texttt{COMPOSITE}. So, for each of those levels there were \(5 \times 3 = 15\) runs per tool. The \texttt{P60} and \texttt{P75} levels have six specification variants (including \texttt{COMPOSITE}) leading to \(6 \times 3 = 18\) runs per tool. Over all five levels, this means there were \(81\) runs per tool. Adding the three baselines, we get \(84\) runs per tool and \(252\) runs in total.

In both systems, the testing tool execution step captures traffic information, and scripts then collect the metrics for our analysis.
Resets are applied before each testing tool execution to reduce cross-run state contamination and carry-over effects, so metric differences can be attributed to the specification variants rather than residual system state.

\subsection{Evaluation Metrics}
\label{Sec:Evaluation Metrics}
We measure the impact of specification faults on testing tool effectiveness using four metric categories:
\begin{enumerate}
    \item \textbf{Code coverage:} Line coverage, Branch coverage.
    \item \textbf{Specification coverage:} Schema Operation coverage, Schema 2xx coverage.
    \item \textbf{Request/response quality:} 2xx request rate, 4xx request rate, 5xx request rate.
    \item \textbf{Behavioral diversity:} Unique Log Templates, Unique Method-Path Pairs.
\end{enumerate}
Code coverage, specification coverage, and 5xx-based fault indicators are the most common metrics used in empirical studies that evaluate~\cite{Golmohammadi2023} and compare~\cite{corradini2024,zhang2023,kim2024} OpenAPI-based black-box testing tools.

We measure code coverage using JaCoCo~\cite{jacoco} and GCOV~\cite{gcovr}. We measure specification coverage from HTTP traces and the OpenAPI specification using Restats~\cite{martinlopez2019coverage,restats}. Schema Operation Coverage reports the percentage of operations that received any response during tool execution. Schema 2xx Coverage reports the percentage of operations that received at least one 2xx response. We separate this from the request/response quality category because Schema 2xx Coverage measures operation-level reachability, whereas the request/response quality metrics capture the request outcome distribution. This helps identify and distinguish scenarios where a tool reaches all declared operations but generates predominantly invalid traffic. We measure the request/response quality metrics from normalized HTTP traces using the standard HTTP status codes. Finally, we measure behavioral diversity using unique log templates, which capture distinct server-side execution patterns~\cite{reinikainen2026assessing}, and unique method-path pairs, which capture the breadth of the HTTP request surface explored by the tool. 

For each metric $m$ in category $C$, we compute the baseline-relative change $\delta_m = v_m^{\text{fault}} - v_m^{\text{baseline}}$, where $v_m^{\text{fault}}$ is the metric value under the faulty specification and $v_m^{\text{baseline}}$ is the value under the baseline specification. We orient each $\delta_m$ so that positive values indicate worse testing behavior. For metrics where higher values indicate better performance (e.g., line coverage, 2xx request rate) we negate the change, while for metrics where higher values indicate worse performance (e.g., 4xx request rate, 5xx request rate) we keep the sign. We orient a higher 5xx as degradation as the service implementation for both systems is held constant across all runs, and only the OpenAPI specification is mutated. The observed 5xx rate thus measures the quality of the requests the tools generate, not the robustness of the system under test. A rise in 5xx means the mutated specification misled the tool into generating poor request traffic that the underlying service could not process. We qualify this as a loss of request/response quality. If the 5xx rate starts falling, we do not read it as an improvement immediately. The RESTler analysis in Section~\ref{RQ2: Tool-Specific Effects of Fault Classes} is one such case, where a lower 5xx rate reflects a broken dependency chain that routes the tool away from failure-prone paths rather than better requests. We thus read the 5xx rate together with the 2xx and 4xx rates and the request/response quality category as a whole. The category degradation score is the mean of the oriented changes within the category:

\begin{equation}
\label{eq:degradation}
D_C = \frac{1}{|C|} \sum_{m \in C} \tilde{\delta}_m,
\end{equation}

\noindent where $\tilde{\delta}_m$ is the oriented baseline-relative change for metric $m$. Concretely, the code coverage degradation score is the mean of the oriented changes in line and branch coverage. The specification coverage degradation score is the mean of the oriented changes in Schema Operation Coverage and Schema 2xx Coverage. The request/response quality degradation score is the mean of the oriented changes in the 2xx, 4xx, and 5xx request rates. The behavioral diversity degradation score is the mean of the oriented changes in unique log templates and unique method-path pairs. The only exception is \texttt{METHOD\_SEM}, where the behavioral diversity score uses the oriented change in unique log templates only. Remapping HTTP methods rewrites method-path pairs by construction, so under this fault the unique method-path pair count reflects the injected relabeling rather than a change in tool exploration. A positive $D_C$ indicates degradation relative to the baseline, a negative $D_C$ indicates improvement.

All degradation scores reported in Section~\ref{sec:Results} (except for the z-standardized heatmap in Figure~\ref{fig:mean_degradation_heatmaps} and the \texttt{METHOD\_SEM} behavioral diversity score) use this formula.

\section{Results}
\label{sec:Results}
\subsection{RQ1: Impact of Fault Classes}
\label{RQ1: Impact of Fault Classes}
Specification faults have a large impact on testing tool behavior. Table~\ref{tab:rq1_pattern_cases} shows that at the highest severity level, the strongest faults degrade individual metric categories by up to 19.2 percentage points for code coverage, 33.3 for specification coverage, 47.9 for request/response quality, and 84.0 for behavioral diversity. The weakest faults, in the same system--tool combinations, produce near zero changes in every category.

\begin{figure*}
    \centering
    \includegraphics[width=0.85\linewidth]{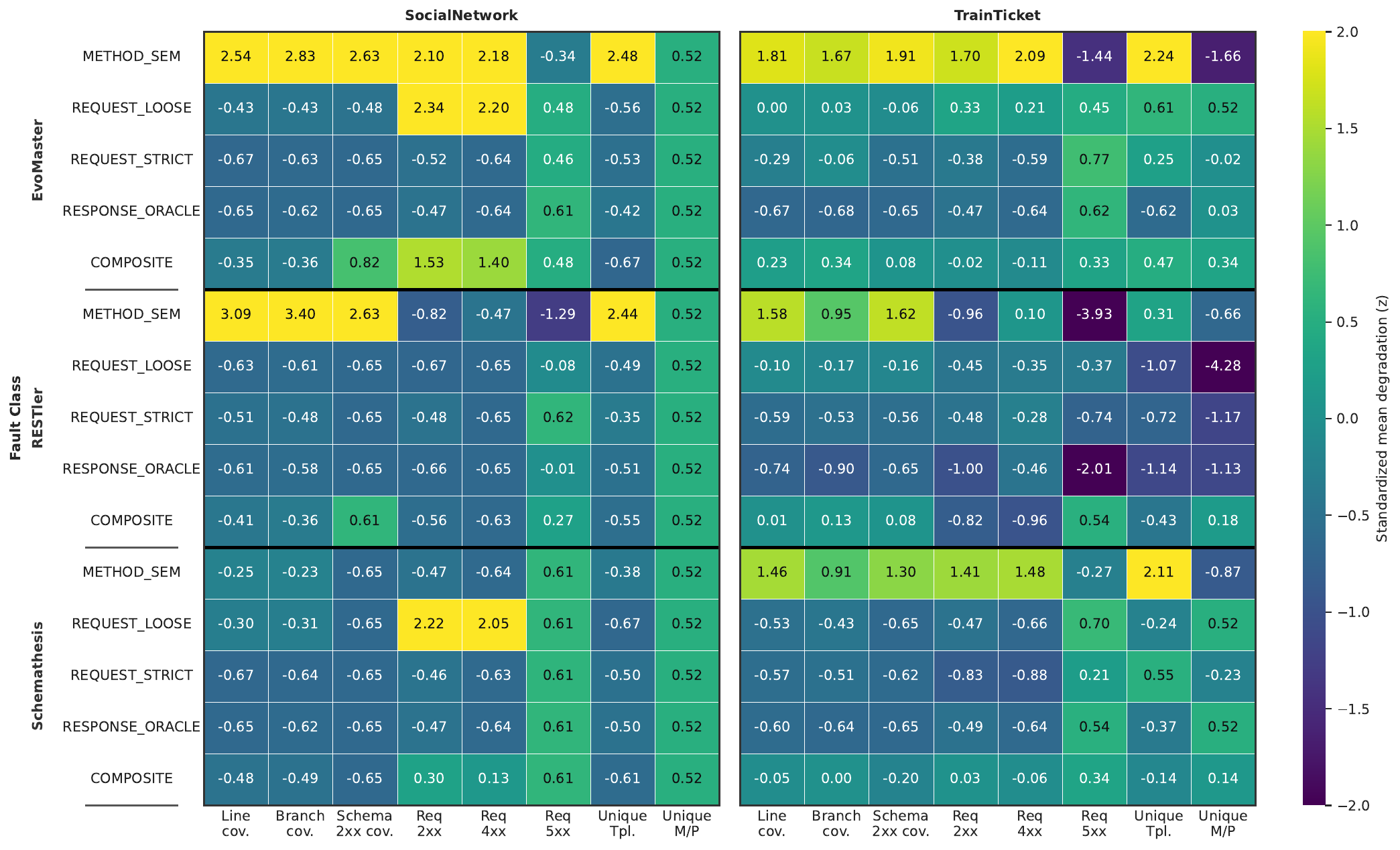}
    \caption{Mean z-standardized degradation heatmaps for the graded fault classes. Each metric column is z-scored globally across all system--tool--fault settings. Positive values mark cells with stronger-than-average degradation on that metric across the full experiment and negative values mark weaker-than-average. Zero indicates average degradation. For SocialNetwork the Unique Method-Path Pairs column stays constant across all fault classes and tools.
    }
    \label{fig:mean_degradation_heatmaps}
\end{figure*}

The injected specification faults produce four distinct degradation patterns across the two systems and three tools. Figure~\ref{fig:mean_degradation_heatmaps} shows the degradation profile of each graded fault across all metrics. We orient the metrics so that higher values always mean worse degradation and z-standardize each metric globally to visualize the fault patterns in a single view. Table~\ref{tab:rq1_pattern_cases} grounds each pattern to raw metric category degradation scores at the highest severity of faults for all system--tool combinations. We also report the observed per-seed range across the three seeds, since a per-seed variance, standard deviation, or confidence interval is not reliable at this sample size. This gives a direct, assumption-free summary of the spread behind each score. The ranges are tight for code and specification coverage and wider for behavioral diversity and some request/response quality cells. These patterns are consistent across all five severity levels (P15--P75) and strengthen as severity increases.

\begin{table*}[ht]
\centering
\caption{All metric category degradation scores for each system--tool--fault scenario at the highest severity. Positive values indicate degradation. Negative values indicate movement below baseline. Zero means no degradation occurred compared to baseline. Each cell also shows the per-seed range across the three seeds in brackets. EM = EvoMaster, RL = RESTler, ST = Schemathesis.}
\label{tab:rq1_pattern_cases}
\setlength{\tabcolsep}{3pt}
\begin{tabular*}{\textwidth}{@{} l l l @{\extracolsep{\fill}} r@{\,}l r@{\,}l r@{\,}l r@{\,}l c @{}}
\hline
\textbf{System} & \textbf{Tool} & \textbf{Fault} & \multicolumn{2}{c}{\textbf{Code Coverage}} & \multicolumn{2}{c}{\textbf{Spec Coverage}} & \multicolumn{2}{c}{\textbf{Req/Resp}} & \multicolumn{2}{c}{\textbf{Diversity}} & \textbf{Pattern} \\
\hline
\multirow[c]{5}{*}{SN} & \multirow[c]{5}{*}{EM} & \texttt{METH\_SEM} & 16.93 & [16.90, 16.95] & 33.33 & [33.33, 33.33] & 38.12 & [36.81, 39.71] & 73.90 & [72.97, 74.36] & Broad degradation \\
 & & \texttt{REQ\_L} & 0.92 & [0.90, 0.95] & 2.78 & [0.00, 8.33] & 47.17 & [46.59, 47.67] & -0.05 & [-2.70, 2.56] & Hidden degradation \\
 & & \texttt{REQ\_S} & -0.37 & [-1.15, 0.05] & 0.00 & [0.00, 0.00] & -0.88 & [-1.40, -0.32] & -0.07 & [-4.05, 3.85] & Flat \\
 & & \texttt{RESP\_O} & 0.00 & [0.00, 0.00] & 0.00 & [0.00, 0.00] & 0.00 & [0.00, 0.00] & 2.14 & [0.00, 3.85] & Flat \\
 & & \texttt{COMP} & 1.27 & [1.05, 1.40] & 11.11 & [8.33, 16.67] & 19.86 & [19.27, 20.46] & -0.50 & [-4.05, 1.28] & Accumulated but diluted \\
\hline
\multirow[c]{5}{*}{SN} & \multirow[c]{5}{*}{RL} & \texttt{METH\_SEM} & 19.15 & [19.15, 19.15] & 33.33 & [33.33, 33.33] & -6.36 & [-9.16, -3.90] & 83.97 & [81.13, 89.66] & Broad degradation \\
 & & \texttt{REQ\_L} & 0.00 & [0.00, 0.00] & 0.00 & [0.00, 0.00] & -4.45 & [-7.14, -2.35] & -0.63 & [-1.89, 0.00] & Flat and reversal-prone \\
 & & \texttt{REQ\_S} & -0.07 & [-0.20, 0.00] & 0.00 & [0.00, 0.00] & -4.82 & [-7.83, -2.53] & -0.31 & [-1.89, 0.94] & Flat and reversal-prone \\
 & & \texttt{RESP\_O} & 0.00 & [0.00, 0.00] & 0.00 & [0.00, 0.00] & -4.90 & [-7.66, -2.59] & -0.63 & [-1.89, 0.00] & Flat and reversal-prone \\
 & & \texttt{COMP} & 0.75 & [0.75, 0.75] & 8.33 & [8.33, 8.33] & -0.07 & [-2.82, 2.22] & 0.00 & [0.00, 0.00] & Accumulated but diluted \\
\hline
\multirow[c]{5}{*}{SN} & \multirow[c]{5}{*}{ST} & \texttt{METH\_SEM} & 1.98 & [1.95, 2.05] & 0.00 & [0.00, 0.00] & 0.00 & [0.00, 0.00] & 2.66 & [2.63, 2.70] & Flat \\
 & & \texttt{REQ\_L} & 0.83 & [0.40, 1.40] & 0.00 & [0.00, 0.00] & 47.90 & [47.24, 48.71] & -1.83 & [-8.11, 1.32] & Hidden degradation \\
 & & \texttt{REQ\_S} & -0.20 & [-0.60, 0.00] & 0.00 & [0.00, 0.00] & 0.20 & [-0.34, 0.70] & -0.01 & [-1.35, 1.32] & Flat \\
 & & \texttt{RESP\_O} & 0.00 & [0.00, 0.00] & 0.00 & [0.00, 0.00] & 0.00 & [0.00, 0.00] & 0.00 & [0.00, 0.00] & Flat \\
 & & \texttt{COMP} & 0.20 & [0.20, 0.20] & 0.00 & [0.00, 0.00] & 5.75 & [5.68, 5.81] & -2.26 & [-8.11, 1.32] & Accumulated but diluted \\
\hline
\multirow[c]{5}{*}{TT} & \multirow[c]{5}{*}{EM} & \texttt{METH\_SEM} & 12.95 & [12.55, 13.45] & 27.25 & [27.01, 27.49] & 32.67 & [32.00, 33.50] & 54.42 & [49.64, 59.15] & Broad degradation \\
 & & \texttt{REQ\_L} & 3.27 & [2.55, 3.65] & 5.37 & [5.21, 5.69] & 13.35 & [12.68, 13.80] & 14.19 & [-7.93, 25.53] & Hidden degradation$^{\dagger}$ \\
 & & \texttt{REQ\_S} & 2.05 & [1.75, 2.45] & 1.66 & [1.42, 1.90] & 1.74 & [1.17, 2.45] & 5.34 & [-13.80, 16.21] & Flat \\
 & & \texttt{RESP\_O} & -0.12 & [-0.65, 0.15] & 0.00 & [0.00, 0.00] & -0.02 & [-0.17, 0.13] & 0.70 & [-14.58, 13.28] & Flat \\
 & & \texttt{COMP} & 5.03 & [4.85, 5.20] & 7.98 & [7.58, 8.29] & 9.71 & [8.99, 10.95] & 15.62 & [6.84, 30.37] & Accumulated but diluted \\
\hline
\multirow[c]{5}{*}{TT} & \multirow[c]{5}{*}{RL} & \texttt{METH\_SEM} & 10.65 & [10.60, 10.75] & 24.88 & [24.88, 24.88] & -3.18 & [-24.22, 11.12] & 28.14 & [26.76, 29.87] & Broad degradation \\
 & & \texttt{REQ\_L} & 2.65 & [2.60, 2.75] & 4.74 & [4.74, 4.74] & 4.76 & [2.45, 7.40] & -56.25 & [-64.48, -48.91] & Reversal-prone \\
 & & \texttt{REQ\_S} & 0.63 & [0.35, 0.80] & 1.11 & [0.95, 1.18] & -2.33 & [-3.18, -1.34] & 2.53 & [-2.11, 6.46] & Flat and reversal-prone \\
 & & \texttt{RESP\_O} & -0.23 & [-0.55, 0.00] & 0.00 & [0.00, 0.00] & -6.15 & [-10.42, -0.82] & -3.51 & [-7.75, 0.00] & Flat and reversal-prone \\
 & & \texttt{COMP} & 4.03 & [3.95, 4.10] & 8.06 & [8.06, 8.06] & 7.80 & [1.66, 12.49] & 10.29 & [7.26, 15.52] & Accumulated but diluted \\
\hline
\multirow[c]{5}{*}{TT} & \multirow[c]{5}{*}{ST} & \texttt{METH\_SEM} & 9.60 & [9.50, 9.80] & 20.93 & [20.85, 21.09] & 30.65 & [30.09, 31.70] & 50.33 & [47.01, 52.38] & Broad degradation \\
 & & \texttt{REQ\_L} & 0.97 & [0.85, 1.05] & 0.08 & [0.00, 0.24] & -0.78 & [-0.92, -0.51] & 5.49 & [3.23, 7.20] & Hidden degradation$^{\ddagger}$ \\
 & & \texttt{REQ\_S} & 0.48 & [0.05, 0.75] & 1.42 & [0.47, 2.13] & -5.15 & [-5.97, -4.01] & -0.04 & [-1.97, 1.79] & Flat and reversal-prone \\
 & & \texttt{RESP\_O} & 0.17 & [-0.15, 0.35] & 0.00 & [0.00, 0.00] & -0.44 & [-0.56, -0.36] & 2.17 & [1.98, 2.38] & Flat \\
 & & \texttt{COMP} & 3.10 & [2.40, 3.50] & 5.06 & [4.50, 5.45] & 9.81 & [9.11, 10.69] & 0.12 & [-0.60, 0.64] & Accumulated but diluted \\
\hline
\multicolumn{12}{l}{\footnotesize $^{\dagger}$ Hidden degradation appears in a milder form, while the gap between coverage and request/response quality is preserved.} \\
\multicolumn{12}{l}{\footnotesize $^{\ddagger}$ Hidden degradation appears in behavioral diversity rather than request/response quality.} \\
\end{tabular*}
\end{table*}

\subsubsection{\textbf{Broad degradation under method-semantic faults}}
\emph{Broad degradation} occurs when all four metric categories degrade together rather than in isolation. Figure~\ref{fig:mean_degradation_heatmaps} shows that \texttt{METHOD\_SEM} causes this pattern. It is the most degradative fault in five of six system--tool combinations. Table~\ref{tab:rq1_pattern_cases} shows highly positive category scores in three of these combinations. In two cases, both with RESTler, the same broad pattern is still visible, but the request/response score reverses. For SocialNetwork with RESTler, code coverage, specification coverage, and diversity all degrade strongly. However, the request/response score turns negative since the drop in 5xx behavior outweighs the rise in 4xx errors. The same reversal appears in milder form for TrainTicket with RESTler. TrainTicket with EvoMaster degrades strongly across all four categories. When method semantics are corrupted, the tools lose the reachability of code paths and specification operations, generate low-quality requests, and alter their exploration behavior.

Broad degradation deepens monotonically with severity in most system--tool combinations. Increasing HTTP operation remapping degrades code coverage, specification coverage, and request/response quality. Under RESTler, request and response quality reverses instead.

\subsubsection{\textbf{Hidden degradation under loosened request constraints}}
\emph{Hidden degradation} occurs when code and specification coverage remain near the baseline while request/response quality or behavioral diversity degrades heavily. Figure~\ref{fig:mean_degradation_heatmaps} shows that \texttt{REQUEST\_LOOSE} causes this pattern. This makes it the most degradative fault in settings where the code and specification coverage metrics remain near baseline. Table~\ref{tab:rq1_pattern_cases} shows \texttt{REQUEST\_LOOSE} producing hidden degradation in four of the six system–tool combinations. The clearest cases are SocialNetwork with EvoMaster and Schemathesis, where code and specification coverage remain close to baseline while request/response degradation reaches 47.17 and 47.90 respectively, exceeding the broad degradation caused by \texttt{METHOD\_SEM}. The same pattern appears in milder form on TrainTicket with EvoMaster, where request/response degradation reaches 13.35 while coverage stays below 5.37. TrainTicket with Schemathesis shows a variant where the degradation appears in behavioral diversity rather than in request/response quality, while coverage metrics again remain near baseline. Relaxing request constraints lets tools generate syntactically accepted but semantically poor traffic, a failure mode invisible to coverage-only evaluation.

The gap between coverage and request/response degradation that defines this pattern widens as severity increases.
\subsubsection{\textbf{Flat and reversal-prone degradation under strict request and response-oracle faults}}
\emph{Flat and reversal-prone} behavior occurs when the overall degradation stays close to zero across all metric category scores, with some scores falling below the baseline. Figure~\ref{fig:mean_degradation_heatmaps} shows that \texttt{REQUEST\_STRICT} and \texttt{RESPONSE\_ORACLE} produce this pattern. Table~\ref{tab:rq1_pattern_cases} confirms that \texttt{RESPONSE\_ORACLE} is flat or reversal-prone in all six system--tool combinations, and \texttt{REQUEST\_STRICT} in five of six. The exception is TrainTicket with EvoMaster, where \texttt{REQUEST\_STRICT} causes mild broad degradation. The strongest metric reversals happen in TrainTicket with RESTler under \texttt{RESPONSE\_ORACLE}, where request/response quality and behavioral diversity scores fall below baseline ($-$6.15 and $-$3.51 respectively). These reversals happen because of two shifts: the 5xx rate falls from 34.6\% to 22.8\%, and the 2xx successful responses rise, indicating that the tool has gotten better at finding faults. However, this is not the case, it is just sending fewer requests down the paths where the faults exist. Overall, tightening constraints or altering response schemas does not substantially change tool behavior. 

These flat and reversal-prone effects remain consistent across severity levels in most system--tool combinations. The exception is RESTler, where the reversal behavior deepens as severity increases.

\subsubsection{\textbf{Accumulated but diluted degradation under composite faults}}
\emph{Accumulated but diluted} degradation occurs when multiple fault effects combine into a measurable but moderate profile. This degradation never overtakes the strongest single-fault degradation in magnitude. Since \texttt{COMPOSITE} is the only fault that injects several fault classes into one specification, it naturally produces this pattern (Figure~\ref{fig:mean_degradation_heatmaps}). At the highest severity in TrainTicket with EvoMaster, Table~\ref{tab:rq1_pattern_cases} shows all four categories positive, reaching 15.62 on behavioral diversity, but no category score reaches the magnitude of \texttt{METHOD\_SEM} or \texttt{REQUEST\_LOOSE}. The same pattern holds across the remaining system--tool combinations. Some component effects of \texttt{COMPOSITE} stay flat or reverse, diluting the overall degradation effect.

To check whether degradation grows with severity, we use the Jonckheere-Terpstra test, a non-parametric trend test that works with three seeds per level. We measure each trend's strength with Kendall's $\tau_b$. Since we run many tests at once, we control the false discovery rate with the Benjamini-Hochberg procedure at $q < 0.05$. A series is a combination of system, tool, fault, and metric category, followed across the five severity levels. This gives $2 \times 3 \times 5 \times 4 = 120$ series, of which 91 are testable. We drop the other 29 as they are either constant across severity or run at only two severity levels, as with the SocialNetwork \texttt{COMPOSITE} case in Section~\ref{Sec:Fault Injection Design}. Out of the 91 testable series, 47 show degradation that grows significantly with severity, with the $\tau_b$ values from $0.42$ to $1.0$ and a median of $0.85$. So the faults differ little in how strongly degradation grows, but differ in how many of their series grow, which lines up with the four patterns. Broad degradation under \texttt{METHOD\_SEM} grows in the most series, 20 of 22. Accumulated degradation under \texttt{COMPOSITE} is next, at 9 of 12. Hidden degradation under \texttt{REQUEST\_LOOSE} grows in 11 of 22. The flat and reversal-prone faults grow in the fewest, with \texttt{REQUEST\_STRICT} significant in only 6 of 21 series and \texttt{RESPONSE\_ORACLE} in only 1 of 14.

\subsubsection{\textbf{RQ1 Answer}}
Specification faults have a large impact in all system--tool combinations and form four distinct degradation patterns. \texttt{METHOD\_SEM} is the most degradative fault class causing broad degradation across all four metric categories in five out of six system--tool combinations. \texttt{REQUEST\_LOOSE} causes hidden degradation where code and specification coverage stay near the baseline but request/response quality degrades heavily. The degradative effects of \texttt{REQUEST\_STRICT} and \texttt{RESPONSE\_ORACLE} remain flat and reversal-prone, with some category scores moving below baseline. \texttt{COMPOSITE} is measurable, but its effect remains below the strongest single-fault case. These four patterns are not artifacts of the highest severity level but are consistent signatures that strengthen as severity increases.

\subsection{RQ2: Tool-Specific Effects of Fault Classes}
\label{RQ2: Tool-Specific Effects of Fault Classes}
\subsubsection{\textbf{Cross-Tool Agreement splits by metric category}}
The tools agree on how faults affect code and specification coverage. Table~\ref{tab:results_tool_agreement} reports pairwise Spearman correlation $\rho$ between each tool pair for every metric, computed over the shared fault--severity cells within each system. For each metric, we rank the shared fault–severity cells by the change each tool observes, then compare the rankings. A high positive $\rho$ means the tools generally agree on which faults have stronger or weaker effects, even if their exact values differ. A value near zero means no clear agreement, while a negative value means the tools tend to order the effects in opposite ways.

Code coverage agreement is consistently high in TrainTicket ($\rho = 0.88$--$0.97$) and ranges from moderate to strong in SocialNetwork ($\rho = 0.36$--$0.83$). Schema 2xx Coverage shows strong agreement where computable ($\rho = 0.73$--$0.99$). Request/response metrics show low agreement across tool pairs ($\rho = -0.34$--$0.66$), and diversity agreement is inconsistent ($\rho = -0.28$--$0.74$). The tools converge on the coverage signal but diverge on the type and quality of traffic they produce from the same faulty specification. For SocialNetwork several of these correlations cannot be computed. With only six operations its Unique Method-Path Pairs count never changes, staying at 6 for EvoMaster, 6 for RESTler and 7 for Schemathesis across every run, and its Schema Operation Coverage stays at 100\%. This leaves 8 of the 27 SocialNetwork tool pairs undefined in Table~\ref{tab:results_tool_agreement}. The split between coverage agreement and traffic disagreement still appears in both systems, because the SocialNetwork request/response and diversity correlations are computable and show the same low agreement. We therefore read the divergence from TrainTicket as the primary system and use SocialNetwork to confirm it.

\begin{table*}[t]
\centering
\caption{Pairwise Spearman rank correlation ($\rho$) between tools across all metrics in the two systems. Each value is computed over the shared fault--severity cells within a system. EM = EvoMaster, RL = RESTler, and ST = Schemathesis.}
\label{tab:results_tool_agreement}
\small
\begin{tabular*}{\textwidth}{l@{\extracolsep{\fill}}cccccc}
\hline
\textbf{Metric}
& \multicolumn{3}{c}{\textbf{TrainTicket}}
& \multicolumn{3}{c}{\textbf{SocialNetwork}} \\
\cline{2-4} \cline{5-7}
& \textbf{EM--RL$\rho$} & \textbf{EM--ST$\rho$} & \textbf{RL--ST$\rho$}
& \textbf{EM--RL$\rho$} & \textbf{EM--ST$\rho$} & \textbf{RL--ST$\rho$} \\
\hline
\textit{Code Coverage} & & \\
Avg Line Coverage          & 0.97 & 0.88 & 0.88 & 0.63 & 0.83 & 0.36 \\
Avg Branch Coverage        & 0.95 & 0.95 & 0.94 & 0.66 & 0.82 & 0.38 \\
\hline
\textit{Specification Coverage} & & \\
Schema Operation Coverage  & n/a$^{\dagger}$ & n/a$^{\dagger}$ & n/a$^{\dagger}$ & n/a$^{\dagger}$ & n/a$^{\dagger}$ & n/a$^{\dagger}$ \\
Schema 2xx Coverage        & 0.99 & 0.73 & 0.75 & 0.93 & n/a$^{\dagger}$ & n/a$^{\dagger}$ \\
\hline
\textit{Request/Response Quality} & & \\
2xx Request Rate           & 0.25 & 0.66 & -0.18 & -0.32 & 0.37 & -0.11 \\
4xx Request Rate           & 0.28 & 0.53 & 0.05 & 0.35 & 0.46 & -0.34 \\
5xx Request Rate           & 0.19 & 0.36 & 0.27 & 0.35 & 0.26 & 0.19 \\
\hline
\textit{Behavioral Diversity} & & \\
Unique Log Templates            & 0.57 & 0.74 & 0.46 & 0.13 & 0.45 & 0.36 \\
Unique Method-Path Pairs        & 0.22 & 0.09 & -0.28 & n/a$^{\dagger}$ & n/a$^{\dagger}$ & n/a$^{\dagger}$ \\
\hline
\multicolumn{7}{l}{\footnotesize $^{\dagger}$ n/a indicates that one or both tools in that pair have a constant series, so $\rho$ cannot be computed.} \\
\end{tabular*}
\end{table*}

\subsubsection{\textbf{Tool-Specific Patterns}}
\label{Tool-Specific Patterns}
The tools diverge in the way the faults affect their request/response and behavioral diversity metrics. The following cases, drawn from Table~\ref{tab:rq1_pattern_cases}, show the largest divergences.

\paragraph{EvoMaster}
EvoMaster diverges most clearly under \texttt{REQUEST\_LOOSE}. In SocialNetwork, Table~\ref{tab:rq1_pattern_cases} shows its request/response degradation score at 47.17, which is close to Schemathesis (47.90) and far from RESTler ($-$4.45). The same hidden degradation appears in TrainTicket, where EvoMaster reaches 13.35 on request/response score while its code and specification coverage stay low at 3.27 and 5.37. Under \texttt{METHOD\_SEM} in TrainTicket, EvoMaster and Schemathesis both show strong request/response degradation (32.67 and 30.65) while RESTler's request/response score instead reverses to $-$3.18.

EvoMaster reaches the declared API surface but generates lower-quality traffic when request-side constraints are loosened. Therefore, its divergence is driven by hidden degradation rather than metric reversal. 

\paragraph{RESTler}
RESTler is the most reversal-prone tool. In SocialNetwork under \texttt{METHOD\_SEM}, Table~\ref{tab:rq1_pattern_cases} shows that its request/response score falls to $-$6.36 while EvoMaster reaches 38.12 and Schemathesis stays flat. In TrainTicket, its strongest reversal appears under \texttt{RESPONSE\_ORACLE}, with a request/response score of $-$6.15. Under \texttt{REQUEST\_LOOSE} in TrainTicket, the strongest cross-tool split appears in diversity, where RESTler drops to $-$56.25 while EvoMaster and Schemathesis show positive degradation (14.19 and 5.49 respectively). This reversal is not unique to SocialNetwork. It appears under three of five fault classes in TrainTicket and four of five in SocialNetwork. This effect is more pervasive in SocialNetwork because its six operations give RESTler shorter dependency chains, so a single broken link reverses its traffic under almost every fault.

These metric reversals follow from RESTler's stateful architecture~\cite{atlidakis2019}. RESTler builds request sequences by extracting dynamic values from 2xx responses and parsing them against the documented response schemas. When \texttt{RESPONSE\_ORACLE} tightens a 2xx schema by adding a required property that the implementation does not return, RESTler's response parser fails to match the schema. This breaks the dependency chain for downstream requests. If those downstream requests trigger 5xx responses in the baseline, their absence reduces the 5xx rate and raises the proportion of successful responses. The tool is not testing more effectively. It sends fewer requests into failure-prone paths as the tightened response schema broke the dependency chain that led to those paths.

\paragraph{Schemathesis}
Schemathesis shows the strongest hidden degradation. In SocialNetwork under \texttt{REQUEST\_LOOSE}, Table~\ref{tab:rq1_pattern_cases} shows that its request/response degradation score reaches 47.90 while code and specification coverage remain near baseline. Under \texttt{METHOD\_SEM} in the same system, it stays flat (0.00 on request/response) while EvoMaster reaches 38.12 and RESTler reverses to $-$6.36. 

This makes Schemathesis the clearest example of a tool where coverage metrics are most misleading. Its coverage stays intact under faults that heavily degrade its traffic quality. The same behavior shows in TrainTicket but through a different metric family. Under \texttt{REQUEST\_LOOSE} the TrainTicket request/response score stays flat at $-$0.78 while behavioral diversity degrades at 5.49. 

\subsubsection{\textbf{RQ2 Answer}}
The three tools converge on how faults affect the code and specification coverage surface but diverge on how faults affect the traffic they generate. We establish this divergence on TrainTicket, where every tool pair can be compared. SocialNetwork confirms this for the request/response quality and behavioral diversity metrics, and shows each divergence more sharply because its six operations concentrate every fault into a sharper signal. EvoMaster shows hidden degradation under \texttt{REQUEST\_LOOSE}, generating lower-quality traffic while maintaining coverage. Schemathesis maintains coverage under faults that heavily degrade its traffic quality. RESTler diverges through metric reversals under \texttt{RESPONSE\_ORACLE} and \texttt{METHOD\_SEM}.

\subsection{RQ3: Degradation signal by metric family}
Coverage metrics capture broad degradation but miss hidden degradation entirely. Table~\ref{tab:rq2_fault_summary} shows which metric families carry signal for each fault using mean category scores per fault across all six system--tool combinations at the highest severity. The score columns show how much degradation is visible in each metric family. 

\begin{table*}[t]
\centering
\caption{Degradation signal by metric family for each fault class. Pattern columns count the faults exhibiting each pattern. Score columns report the mean degradation score per metric family, averaged across the six system--tool combinations.}
\label{tab:rq2_fault_summary}
\small
\begin{tabular}{lcccc c cccc} 
\hline
\textbf{Fault}
& \multicolumn{4}{c}{\textbf{Patterns}}
& 
& \multicolumn{4}{c}{\textbf{Mean Degradation Score}} \\
\cline{2-5} \cline{7-10}
& \textbf{Broad} & \textbf{Hidden} & \textbf{Flat/Reversal} & \textbf{Accum.}
& 
& \textbf{Code Coverage} & \textbf{Spec Coverage} & \textbf{Request/Response} & \textbf{Diversity} \\
\hline
\texttt{METH\_SEM} & 5 & 0 & 1 & 0 && 11.88 & 23.29 & 15.32 & 48.90 \\
\texttt{REQ\_L}    & 0 & 4 & 2 & 0 && 1.44  & 2.16  & 17.99 & $-$6.51 \\
\texttt{REQ\_S}    & 0 & 0 & 6 & 0 && 0.42  & 0.70  & $-$1.87 & 1.24 \\
\texttt{RESP\_O}   & 0 & 0 & 6 & 0 && $-$0.03 & 0.00 & $-$1.92 & 0.14 \\
\texttt{COMP}      & 0 & 0 & 0 & 6 && 2.40  & 6.76  & 8.81  & 3.88 \\
\hline
\multicolumn{7}{l}{\footnotesize Per-system values appear in Table~\ref{tab:rq1_pattern_cases}} \\
\end{tabular}
\end{table*}

\texttt{METHOD\_SEM} shows positive values across all four metric category scores showing that all four metric categories carry the signal required to identify its effects. 
Behavioral diversity carries the strongest signal (48.90, from log templates only), followed by specification coverage (23.29), request/response quality (15.32), and code coverage (11.88).

\texttt{REQUEST\_LOOSE} is where coverage metrics alone fail. Code and specification coverage degradation scores average 1.44 and 2.16 while request/response reaches 17.99. A coverage-only evaluation would classify the effects of this fault as weak or not degradative at all despite it producing the highest request/response degradation of any fault class. Hidden degradation does not register in coverage metrics.

\texttt{REQUEST\_STRICT} and \texttt{RESPONSE\_ORACLE} show little degradation in any metric category, with negative request/response scores indicating reversals ($-$1.87 and $-$1.92). Both faults produce near-zero scores across all four columns, leaving no degradation for any metric to detect.

\texttt{COMPOSITE} shows moderate scores across all columns, with the strongest signal in request/response (8.81) and specification coverage (6.76). No single metric family captures the full composite effect.

\subsubsection{\textbf{RQ3 Answer}}
Coverage alone cannot reliably distinguish a well-functioning tool from one that is reaching endpoints but generating rejected requests. Coverage detects \texttt{METHOD\_SEM} and partially detects \texttt{COMPOSITE}, but misses \texttt{REQUEST\_LOOSE} entirely. Request/response is the only metric family that detects hidden degradation. Treating coverage as the sole evaluation metric can hinder the visibility of some hidden fault effects.

\section{Threats to Validity}
\label{sec:ThreatstoValidity}

A threat to construct validity is that our fault taxonomy may not cover all possible specification faults. It covers common issues in OpenAPI-based API testing, while semantic faults involving inter-operation dependencies or business-logic constraints fall outside our six fault classes. This is by construction: an OpenAPI specification describes each operation independently and cannot express ordering, shared state, inter-operation data dependencies, or business-logic constraints. Thus, faults in these domains cannot be represented by mutating an OpenAPI specification. Within OpenAPI's expressive scope, each class is grounded in documented fault types from prior literature, as cited per class in Table~\ref{tab:faultclass-grounding} and motivated in Section~\ref{sec:Fault taxonomies and mutation-testing background}. An example is the HTTP verb remapping in \texttt{METHOD\_SEM}, a REST antipattern that departs from stable HTTP method conventions. We inject it in its strongest form, fully remapping the method of each affected operation, to characterize the upper bound of method-semantic degradation. Milder real-world variants of verb misuse would degrade testing less at the same severity, so our results show the largest impact this fault class can have, not how often it occurs. We use metrics that are common in OpenAPI-based black-box testing. For instance, HTTP status codes are used as a proxy for behavioral correctness and unique log templates for server-side execution diversity. These proxies might be imperfect.

A threat to internal validity concerns the manual construction of our baseline OpenAPI specifications. This process could have introduced specification errors. However, we use baseline-relative values for analysis, and our baseline remains constant. This mitigates any errors already present in the baseline specifications. We also control for tool randomness using fixed seeds and strict reset isolation between runs. This ensures that every run starts from its own deployment and environment, without cross-contamination from previous runs.

A potential threat to external validity lies in our choice of systems and tools. We evaluate a large benchmark MSS  (TrainTicket: 211 operations) and a small one (SocialNetwork: 6 operations), and three testing tools based on distinct black-box strategies: search-based (EvoMaster), stateful fuzzing (RESTler), and property-based (Schemathesis). The small operation count in SocialNetwork limits the generalizability of its estimates, so we treat it as a confirmatory case. The RQ1 degradation patterns generalize across both systems. The RQ2 divergence, however, rests primarily on TrainTicket, although the divergence in request/response and behavioral diversity does hold in both.
Our tools do not cover learning-based or reinforcement-learning testing tools, and the metric reversals we observe for RESTler are tied to its stateful dependency-chain architecture. Expanding the evaluation to additional MSS and tool families would strengthen these findings.
\section{Conclusion}
\label{sec:Conclusion}
In this paper, we studied how faults in OpenAPI specifications affect the effectiveness of black-box testing tools in MSS. We defined a taxonomy of six OpenAPI specification fault classes grounded in prior literature. We injected these faults at five severity levels into the baseline specifications of two MSS benchmarks, TrainTicket as our primary system and SocialNetwork as a smaller confirmatory case. We ran three popular black-box testing tools (EvoMaster, RESTler, and Schemathesis) on the faulty specifications. We measured the impact using code coverage, specification coverage, request/response quality and behavioral diversity.

We found that specification faults affect testing tool behavior in every system--tool combination. We identified four degradation patterns across the fault classes. Corrupting the HTTP method semantics caused broad degradation across all metrics. Loosening the request constraints caused hidden degradation where code and specification coverage stayed flat but request/response quality degraded heavily. Other faults caused weak effects and some even led to metric reversals. Higher severity amplified each pattern without changing its effect type. The three tools agreed on how faults affect coverage, but diverged on the traffic they generated. This happened notably in EvoMaster and Schemathesis through hidden degradation and in RESTler through metric reversals. These divergences stayed invisible in the coverage-only metrics and only became clear once request/response quality and behavioral diversity metrics were considered. 

Therefore, we conclude that specification quality is a critical factor in OpenAPI-based black-box API testing effectiveness, and coverage metrics alone are not enough to detect it. Evaluations of black-box API testing tools should include request/response quality and behavioral diversity alongside coverage, and practitioners should treat the OpenAPI specification as a testable artifact rather than a fixed input.

The taxonomy in this paper gives practitioners a practical way to categorize specification faults and reason about their impact. Our severity results show that the most degrading faults, \texttt{METHOD\_SEM} and \texttt{REQUEST\_LOOSE}, can cause substantial hidden degradation while not being well captured by standard specification checks. This may help practitioners detect otherwise unnoticed faults, prioritize the fault classes that matter most, and choose appropriate repair strategies. Such faults can be detected by tracking request quality, response quality, and behavioral diversity against a baseline, since these are the metric families that reveal the hidden degradation that coverage misses. A drop in these families while coverage stays near baseline points to relaxed request constraints, whereas a broad drop across all families points to method-semantic faults. Constraint-related classes such as \texttt{REQUEST\_LOOSE} and \texttt{REQUEST\_STRICT} are natural candidates for the specification enrichment approaches discussed in Section~\ref{subsec:SpecificationEnrichment}. Semantic faults such as \texttt{METHOD\_SEM} are not addressed by constraint enrichment and must be fixed at the specification source, where antipattern detection approaches can locate them~\cite{10.1007/978-3-662-45391-9_16,10.1007/978-3-319-46295-0_10}.

For future work, we aim to validate our findings with other MSS where the OpenAPI specifications are created manually. The baseline OpenAPI specifications for both TrainTicket and SocialNetwork, the injector scripts with their injection catalogs, and the mutated specifications are available in our replication package: \url{https://doi.org/10.5281/zenodo.21342161}.

\section*{Acknowledgment}
This work was part of Finland's Ministry of Education and Culture's Doctoral Education Pilot under Decision No.\ VN/3137/2024-OKM-6 (The Finnish Doctoral Program Network in Artificial Intelligence, AI-DOC). This work was supported by the Research Council of Finland (Grant id: 359861). The authors acknowledge CSC-IT Center for Science, Finland, for providing the necessary computing resources.

\bibliographystyle{IEEEtran}
\bibliography{compressed}

\end{document}